\documentclass[lettersize,journal]{IEEEtran}
\usepackage{amsmath,amsfonts}
\usepackage{algorithmic}
\usepackage{algorithm}
\usepackage{array}
\usepackage[caption=false,font=normalsize,labelfont=sf,textfont=sf]{subfig}
\usepackage{textcomp}
\usepackage{stfloats}
\usepackage{url}
\usepackage{verbatim}
\usepackage{graphicx}
\usepackage{cite}

\usepackage{braket}

\begin{document}

\title{Spatiotemporal Multiplexed Rydberg Receiver}

\author{Samuel H. Knarr, Victor G. Bucklew,  Jerrod Langston,  Kevin C. Cox, Joshua C. Hill, David H. Meyer, James A. Drakes
\thanks{Victor G. Bucklew and Samuel H. Knarr are co-first authors. (Corresponding authors: Samuel H. Knarr; Victor G. Bucklew.)}% <-this % stops a space
\thanks{Samuel H Knarr, Victor G. Buckew, Jerrod Langston, and James A. Drakes are with L3Harris Technologies Inc., Palm Bay, FL 32905 USA.}%
\thanks{Kevin C. Cox, Joshua C. Hill, and David H. Meyer are with DEVCOM Army Research Laboratory, Adelphi MD 20783 USA.}}%



\maketitle

\begin{abstract}
Rydberg states of alkali atoms, where the outer valence electron is excited to high principal quantum numbers, have large electric dipole moments allowing them to be used as sensitive, wideband, electric field sensors. These sensors use electromagnetically induced transparency (EIT) to measure incident electric fields. The characteristic timescale necessary to establish EIT determines the effective speed at which the atoms respond to time-varying RF radiation. Previous studies have predicted that this EIT relaxation rate causes a performance roll-off in EIT-based sensors beginning at a less than 10 MHz RF data symbol rate. Here, we propose an architecture for increasing the response speed of Rydberg sensors to greater than 100 MHz, through spatio-temporal multiplexing (STM) of the probe laser. We present experimental results validating the architecture's temporal multiplexing component using a pulsed laser. We benchmark a numerical model of the sensor to this experimental data and use the model to predict the STM sensor's performance as an RF communications receiver. For an on-off keyed (OOK) waveform, we use the numerical model to predict bit-error-ratios (BERs) as a function of RF power and data rates demonstrating feasibility of error free communications up to 100 Mbps with an STM Rydberg sensor. 	
\end{abstract}


\section{Introduction}
\IEEEPARstart{R}{adio} frequency (RF) communication and sensing systems are tasked with supporting increasingly complex applications. There is strong demand for capabilities utilizing new frequency bands, more complex waveforms, and increasingly higher data rates. As these requirements have evolved, new paradigms for RF sensing have emerged as potential alternatives to traditional narrow-band RF antenna receivers. Receivers using Rydberg atoms \cite{lukin2003colloquium} are an emerging platform for highly sensitive detection of incident electric fields over a wide range of frequencies, from dc to THz \cite{jing2020atomic,meyer2021waveguide,holloway2014broadband,anderson2016optical,holloway2017electric,meyer2020assessment}. Rydberg receivers have even been used to demonstrate SI-traceable calibration standards of RF fields \cite{holloway2014broadband}. In this area, Rydberg receivers have a clear advantage over current methods. 

A common two-laser approach for generating Rydberg-electromagnetically induced transparency (EIT) uses narrowband, continuous wave (CW), optical probe and coupling beams. This two-photon interaction creates a transparency window depending on the probe beam's frequency by establishing a coherent superposition of the ground and Rydberg states \cite{fleischhauer2005electromagnetically}. This creates a so-called "dark-state" that is transparent to the probe beam. Within this scheme, incoming RF radiation directly modulates the probe beam transmission. Previous studies into the quantum-limited performance of CW based Rydberg sensors show a  performance roll-off that begins at a less than 10 MHz symbol rate \cite{cox2018quantum}. 

These CW approaches use the EIT steady-state response, such that changes in the probe beam can only be measured when atoms reset into the EIT dark state. This EIT relaxation rate causes a bandwidth limitation for detecting time-varying RF fields. The poor performance of these CW Rydberg sensors at data symbol rates larger than 10 MHz narrows the potential application space of Rydberg sensors. 

In this paper, we overcome this limit and introduce a receiver concept for bridging this performance gap based on spatiotemporal shaping of the probe beam within a Rydberg vapor cell. Our architecture splits the probe beam into spatially separated pulse trains that are temporally delayed relative to each other. Each beam travels through a different physical area of the vapor cell and samples the RF field at a different time. This excitation exploits the asymmetric transients in the EIT turn-on and turn-off times in the atoms\cite{cox2018quantum,bohaichuk2022origins}. The sensor's response time is then set by the transient behavior in the EIT turn-on time. The spatio-temporal multiplexing (STM) architecture allows us to exploit these fast transients repeatedly by avoiding hysteretic effects caused by the EIT relaxation dynamics. By appropriately interleaving these probe pulses, the STM architecture continuously samples the RF field. We numerically simulate the atomic time dynamics to show sampling rates exceeding 100 MHz are possible for RF waveforms with powers in the interaction region of $-30$ to $-40$ dBm. We also benchmark our simulations to experimental results showing good agreement for an RF field of approximately $0.6$ V/m, validating our method.

Throughout the rest of this paper, we introduce and validate the STM-receiver concept. First, we define the architecture of an STM-receiver. Next, we present experimental results detailing the response of a Rydberg sensor to probe pulses of varying width. Finally, we use experimental results to benchmark simulations of the STM-receiver, extracting bit-error-ratios (BERs) as a function of RF power and data rate.

\section{Spatiotemporal Multiplexed Rydberg Receiver}
\begin{figure}[!h]
	\centering
	\includegraphics[width=2in]{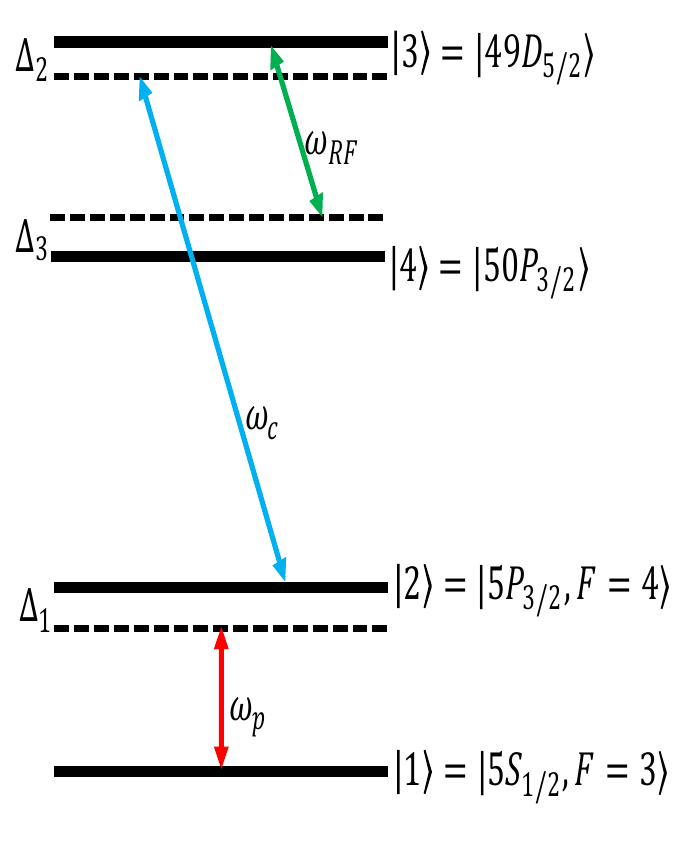}
	\caption{Rubidium energy levels used in this work. We define each beam's detuning (dashed lines) from resonance (solid lines) as $\Delta=\omega_{ij}-\omega_k$ where $\omega_{ij}$ is the transition frequency between states $i$ and $j$, and $\omega_k$ is the beam's optical frequency.}
	\label{fig_1}
\end{figure}
%\Figure[h!](topskip=0pt, botskip=0pt, midskip=0pt)[width=2in]{Figures/Knarr1.pdf}
%{Rubidium energy levels used in this work. We define each beam's detuning (dashed lines) from resonance (solid lines) as $\Delta=\omega_{ij}-\omega_k$ where $\omega_{ij}$ is the transition frequency between states $i$ and $j$, and $\omega_k$ is the beam's optical frequency.\label{fig_1}}

The relevant Rubidium 85 states for the Rydberg-EIT scheme used in this work are shown in Fig \ref{fig_1}. The probe beam with frequency $\omega_p$, excites the atoms from the ground state $\ket{1}$ to the intermediate state $\ket{2}$. From there, a coupling beam with frequency $\omega_c$ drives the atom into a Rydberg state $\ket{3}$ and opens an EIT window for the probe beam in a narrowband region around this two-photon transition. In order to partially cancel Doppler shifts, the coupling beam counter-propagates relative to the probe beam. By sweeping the detuning of either the probe ($\Delta_1$) or coupling ($\Delta_2$) beam and monitoring the probe beam transition, we can map out the EIT transmission window (see Fig \ref{fig_4}). The presence of an RF signal that couples $\ket{3}$ to another Rydberg level $\ket{4}$, causes an Autler-Townes (AT) splitting of the EIT peak. From this splitting the amplitude of the RF field can be determined. See the Appendix for a full theoretical description of this system. As mentioned above, this interrogation scheme typically uses frequency-swept narrowband CW probe and coupling beams and has been used to measure amplitude-modulated (AM), frequency-modulated (FM), and phase-modulated RF signals \cite{meyer2018digital,holloway2019detecting}.

The instantaneous bandwidth (IBW) of an RF field sensor determines the sensor's sampling speed of the RF E-field. For a CW-based Rydberg sensor, this is limited by the EIT relaxation times as atoms repopulate the EIT dark state. CW-based schemes in the literature report a roll-off in signal-to-noise ratio (SNR) for sampling rates less than 10 MHz \cite{meyer2018digital,cox2018quantum}. However, this is much slower than the EIT turn-on transients and Rydberg-Rydberg population transfer in $\ket{3}\rightarrow\ket{4}$ \cite{sapiro2020time,bohaichuk2022origins}. 

%\Figure[h!](topskip=0pt, botskip=0pt, midskip=0pt)[width=3in]{Figures/Knarr2.png}
%{Diagram of a 4-beam version of a spatiotemporal Rydberg receiver for increasing the sampling rate of Rydberg atom based electric field sensors. CW: CW laser. Pulse Shaper: EOM or AOM for pulsing the probe laser. $\tau_n$: temporal delay element for temporally multiplexing the spatially multiplexed probe beams.  PD: photodiode. \label{fig_2}}

\begin{figure}[!h]
	\centering
	\includegraphics[width=3in]{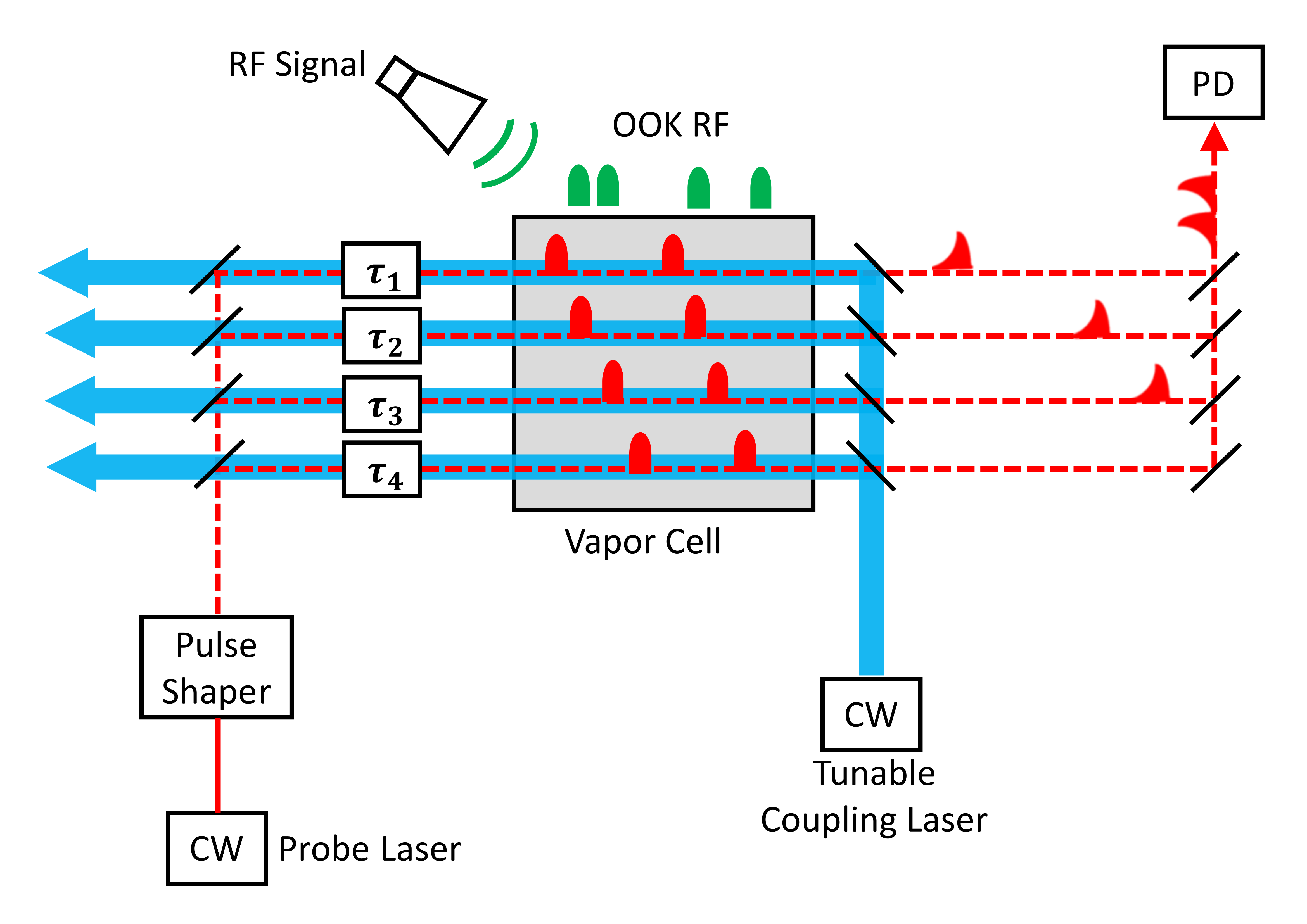}
	\caption{Diagram of a 4-beam version of a spatiotemporal Rydberg receiver for increasing the sampling rate of Rydberg atom based electric field sensors. CW: CW laser. Pulse Shaper: EOM or AOM for pulsing the probe laser. $\tau_n$: temporal delay element for temporally multiplexing the spatially multiplexed probe beams.  PD: photodiode.}
	\label{fig_2}
\end{figure}

Fig \ref{fig_2} shows a conceptual diagram of our solution to this bandwidth limitation: the STM Rydberg receiver. In an STM receiver, the CW probe laser propagates through a pulse shaper, such as an acousto-optic or electro-optic modulator, producing a train of probe pulses. This pulse train is then split into distinct spatial paths through the vapor cell. A CW coupling beam counter-propagates with each probe beam individually or as one large beam in the vapor cell. After propagating through the cell, the probe beams are interleaved and collected on a detector. Depending on requirements, either an array of detectors or a single detector could be used. The STM receiver is also amenable to an RF local oscillator techniques. Below we show how the STM can be used to increase the sampling rates of the Rydberg receivers over their CW counterparts and achieve continuous sampling of the RF field. 

First, we set the repetition rate of the pulse shaper such that the atoms can relax and reach thermal equilibrium between probe pulses. We then set the pulse widths to be shorter than the EIT relaxation times, where atomic populations have not yet reached steady state, but where they display measurable transient features. This allows for sampling the E-field at the pulse bandwidth.  Transients in the EIT establishment have been shown to occur on time scales as short as 10 ns, enabling sampling rates greater than 100 MHz \cite{sapiro2020time}.

We delay each beam relative to the other beams in integer multiples of the pulse width. If we assume that only one probe pulse is needed per RF bit, then the sampling rate, $f_s$, is set by the inverse of the probe pulse width $T$. The number of beams required to achieve continuous sampling of the RF field is set by the repetition rate, $f_r$, and the desired sampling rate, and is given by $N=f_s/f_r$. 

To summarize the STM architecture features, pulsing the probe beam allows sampling of the RF field at rates determined by the EIT transients. Spatially splitting and staggering these pulses, ensures continuous sampling of the RF source. Using both spatial and temporal multiplexing in this way increases the response speed of Rydberg receivers.

\section{Pulsed Probe Experiment}
%\Figure[h!](topskip=0pt, botskip=0pt, midskip=0pt)[width=3in]{Figures/Knarr3.pdf}
%{Schematic of the experimental layout. CW-p: CW probe laser. CW-c: CW coupling laser. AOM: acousto-optic modulator. EOM: electro-optic modulator. DM: dichroic mirror. PD: photodiode. T: Bias-Tee \label{fig_3}}

\begin{figure}[!h]
	\centering
	\includegraphics[width=3in]{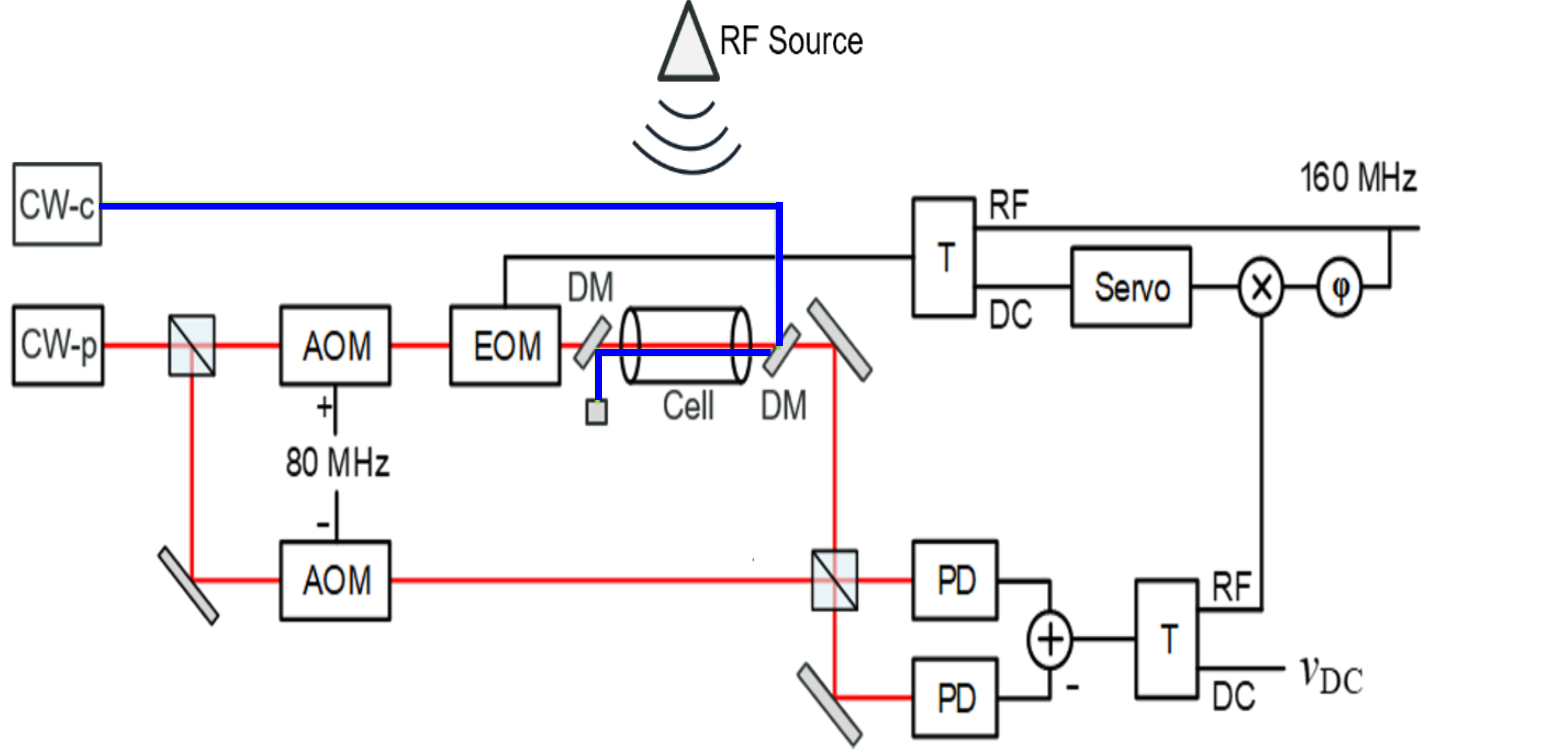}
	\caption{Schematic of the experimental layout. CW-p: CW probe laser. CW-c: CW coupling laser. AOM: acousto-optic modulator. EOM: electro-optic modulator. DM: dichroic mirror. PD: photodiode. T: Bias-Tee}
	\label{fig_3}
\end{figure}

The performance of the proposed STM-receiver depends on the response of a Rydberg cell to pulsed probe beams. The response of Rydberg-atom-based sensors to pulsed RF sources, and pulsed coupling lasers, have been previously investigated and  demonstrated \cite{bohaichuk2022origins,sapiro2020time}. 

Fig 3 shows the experimental layout for measuring a Rubidium vapor cell's response to probe pulses of varying widths. Specifically, we measure the transmitted probe laser power with the RF radiation on and off, as a function of probe pulse width. The coupling laser is on continuously, establishing EIT with the probe pulses. We measure the probe transmission through the vapor cell with a balanced optical homodyne technique. As shown in Fig 4 of a similar setup in Rb 87, the AT splitting of the EIT peak in the presence of RF radiation reduces the measured probe power around the zero probe detuning. The difference in detected probe power at this frequency provides a metric for assessing the performance of this RF receiver.

%\Figure[h!](topskip=0pt, botskip=0pt, midskip=0pt)[width=3in]{Figures/Knarr4}
%{Experimentally measured probe laser transmission as a function of center frequency when an 18 GHz RF source is on, and when it is off in Rb 87. When an RF field is off, EIT is observed, and when an RF field is on, AT splitting of the EIT peak is observed. The secondary peak around 60 MHz is an EIT peak from another nearby Rydberg state. In the STM receiver, the probe and coupling beams are locked to the zero detuning. \label{fig_4}}

\begin{figure}[!h]
	\centering
	\includegraphics[width=3in]{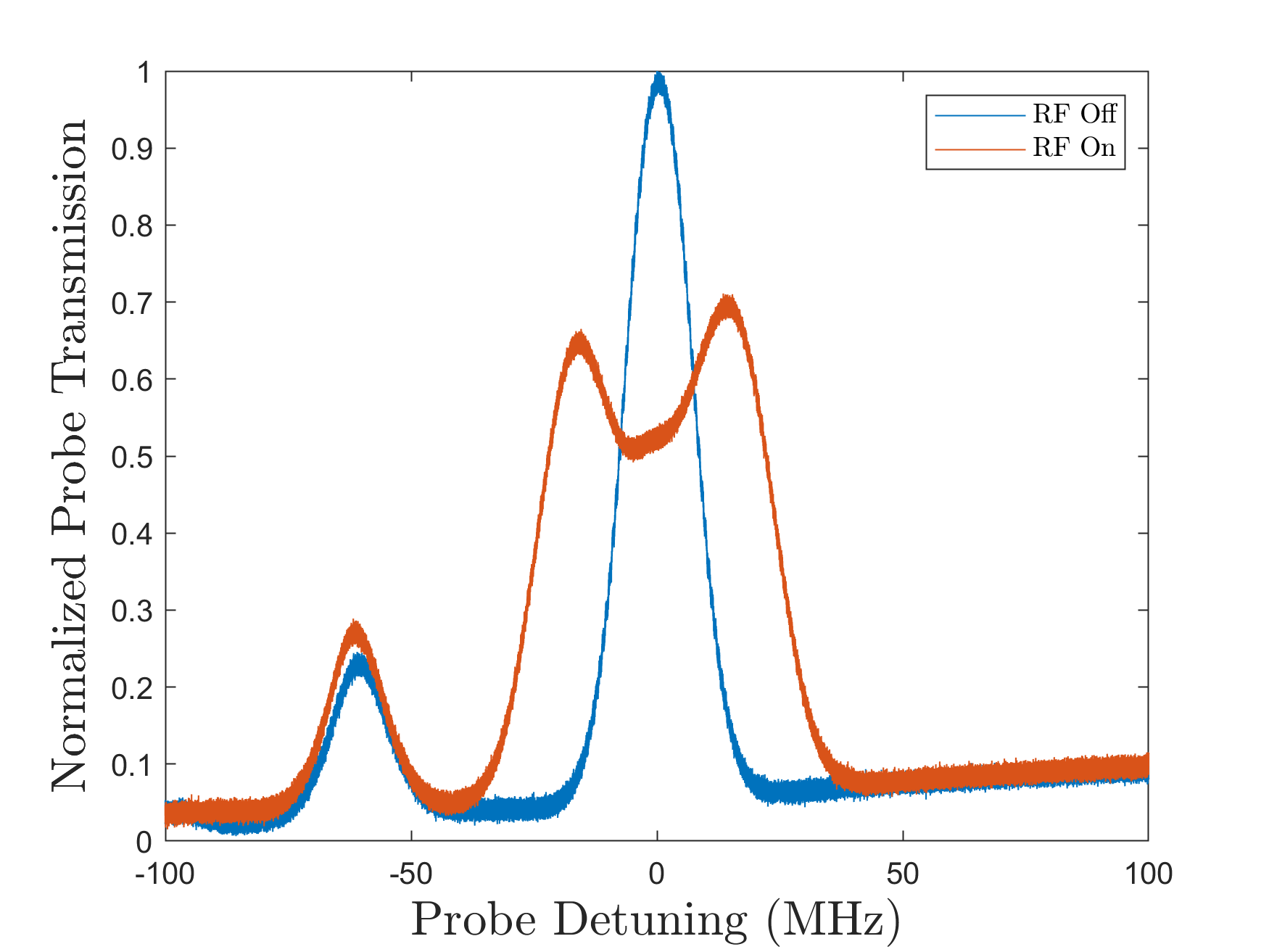}
	\caption{Experimentally measured probe laser transmission as a function of center frequency when an 18 GHz RF source is on, and when it is off in Rb 87. When an RF field is off, EIT is observed, and when an RF field is on, AT splitting of the EIT peak is observed. The secondary peak around 60 MHz is an EIT peak from another nearby Rydberg state. In the STM receiver, the probe and coupling beams are locked to the zero detuning.}
	\label{fig_4}
\end{figure}

Our vapor cell contains naturally abundant Rb held at room temperature. The probe beam is locked to the $\ket{1}=\ket{5S_{1/2},F=3}\rightarrow\ket{2}=\ket{5P_{3/2},F=4}$ transition, has a $e^{-2}$  spot size of 380 $\mu$m, and has 4.7 $\mu$W power before entering the vapor cell. The counter-propagating coupling beam is cavity-locked to the $\ket{2}\rightarrow\ket{3}=\ket{49D_{5/2}}$ transition, has a $e^{-2}$  spot size of 400 $\mu$m, and 600 mW power. An 18.14 GHz RF source at 0 dBm drives the transition $\ket{3}\rightarrow\ket{4}=\ket{50P_{3/2}}$. All fields are co-polarized vertically with respect to the optical table. The balanced detector is configured for homodyne detection in the amplitude quadrant. More detailed description of the experimental architecture can be found in \cite{meyer2020assessment}. By sweeping the coupling beam's frequency through the two-photon EIT resonance in the presence of an RF field we measure the AT splitting of the EIT spectrum to determine the Rabi frequency of the RF beam for benchmarking numerical results ($\Omega_{RF}= 2\pi \times 17.11$ MHz). 

We simulate the propagation of the optical probe beam through the Rydberg cell by solving a time-dependent 4-level model of the atomic populations. This coupled system of linear differential equations is solved with a fourth order Runge-Kutta approach. We use the results of the 4-level model to determine the probe beam's time dependent absorption coefficient through the vapor cell. Shot noise, detector dark and thermal noise, and the spectral response profile of the balanced photodiode are included in the simulation. To match the scale of the experimental results, we scale our numerical results by a factor of order one. Details on the numerical model are included in the Appendix.

\begin{figure*}[!t]
	\centering
	\subfloat[]{\includegraphics[width=2.5in]{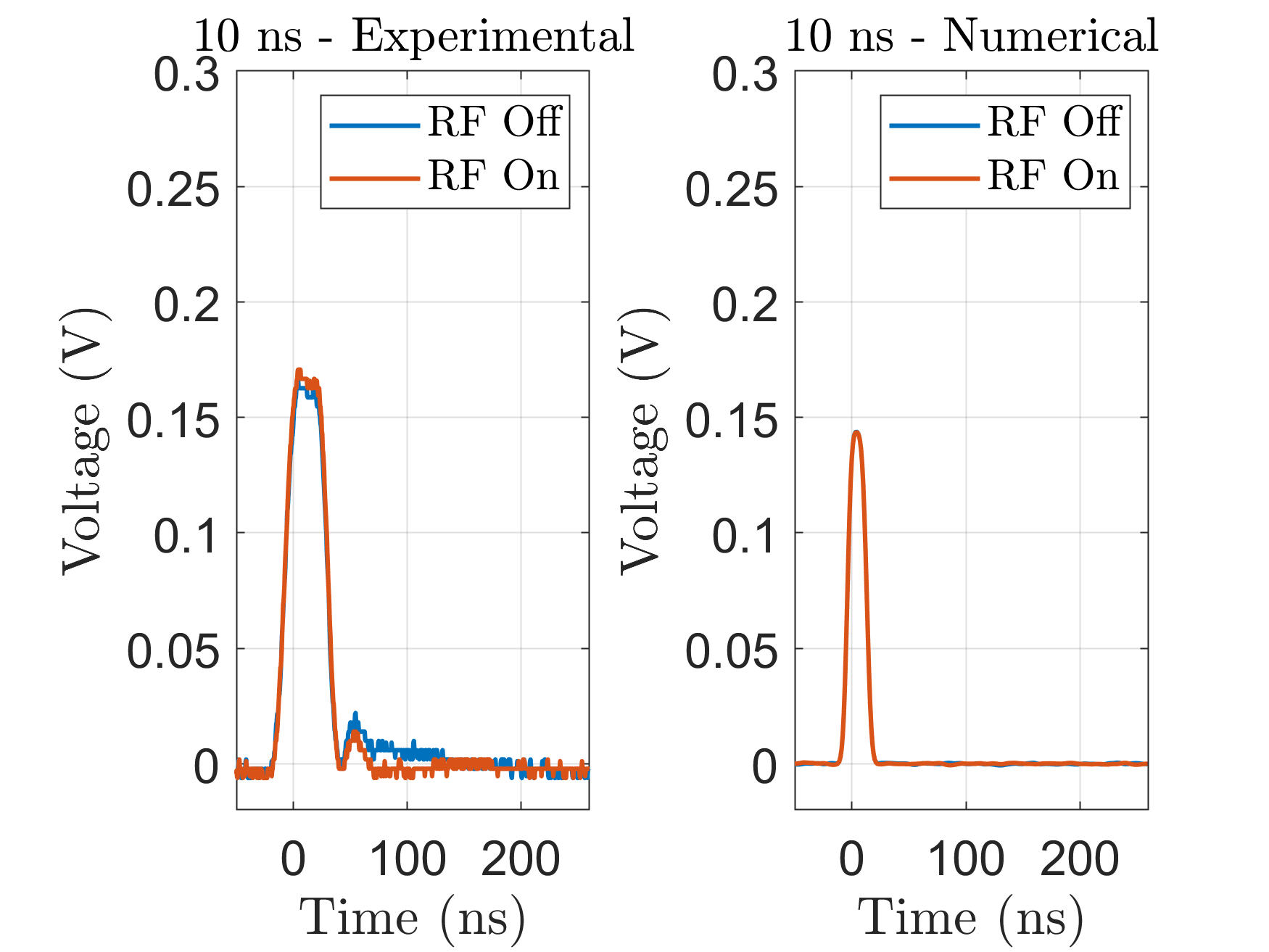}
		\label{10ns}}
	\subfloat[]{\includegraphics[width=2.5in]{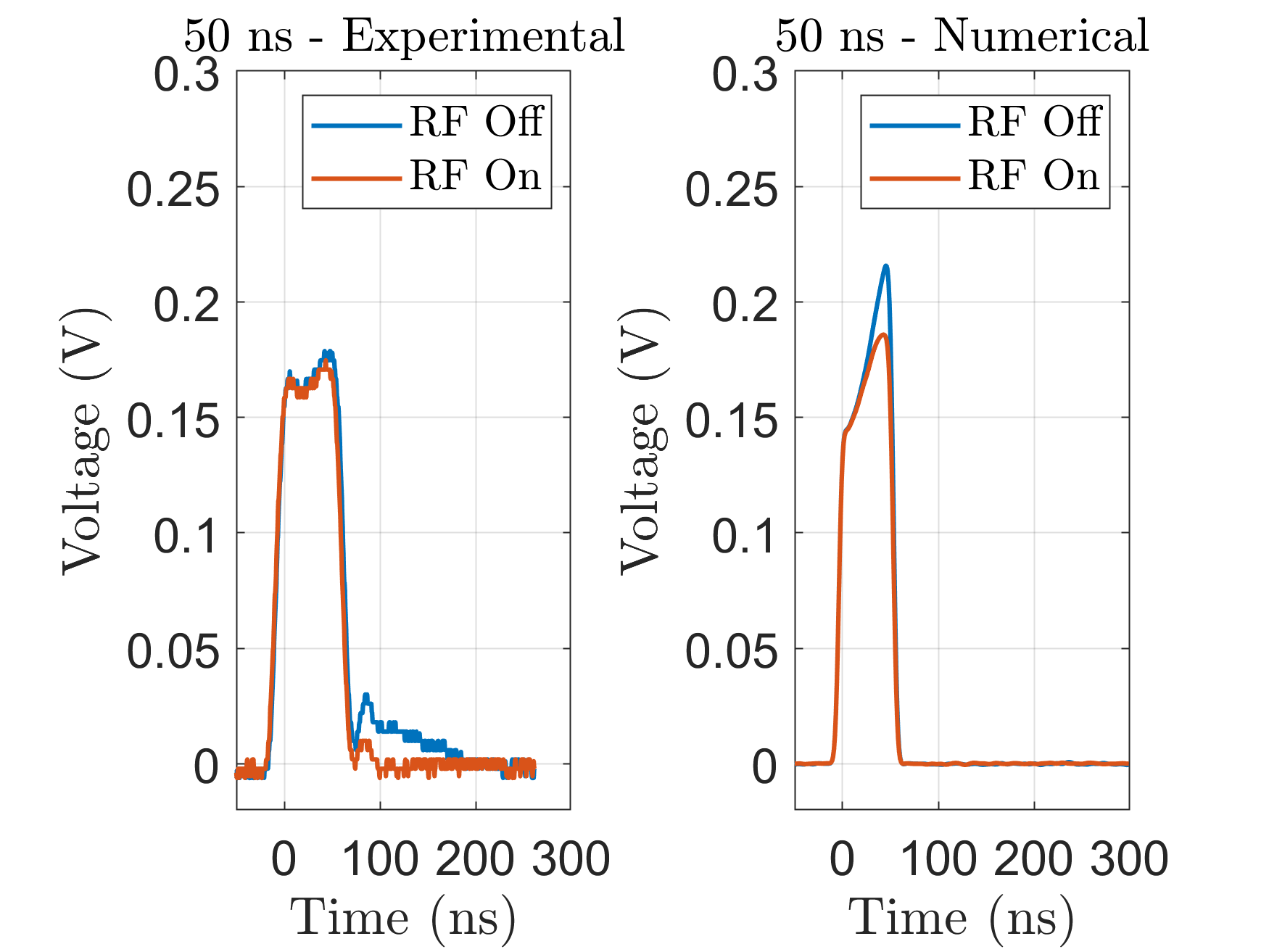}
		\label{50ns}}\hfil
	\subfloat[]{\includegraphics[width=2.5in]{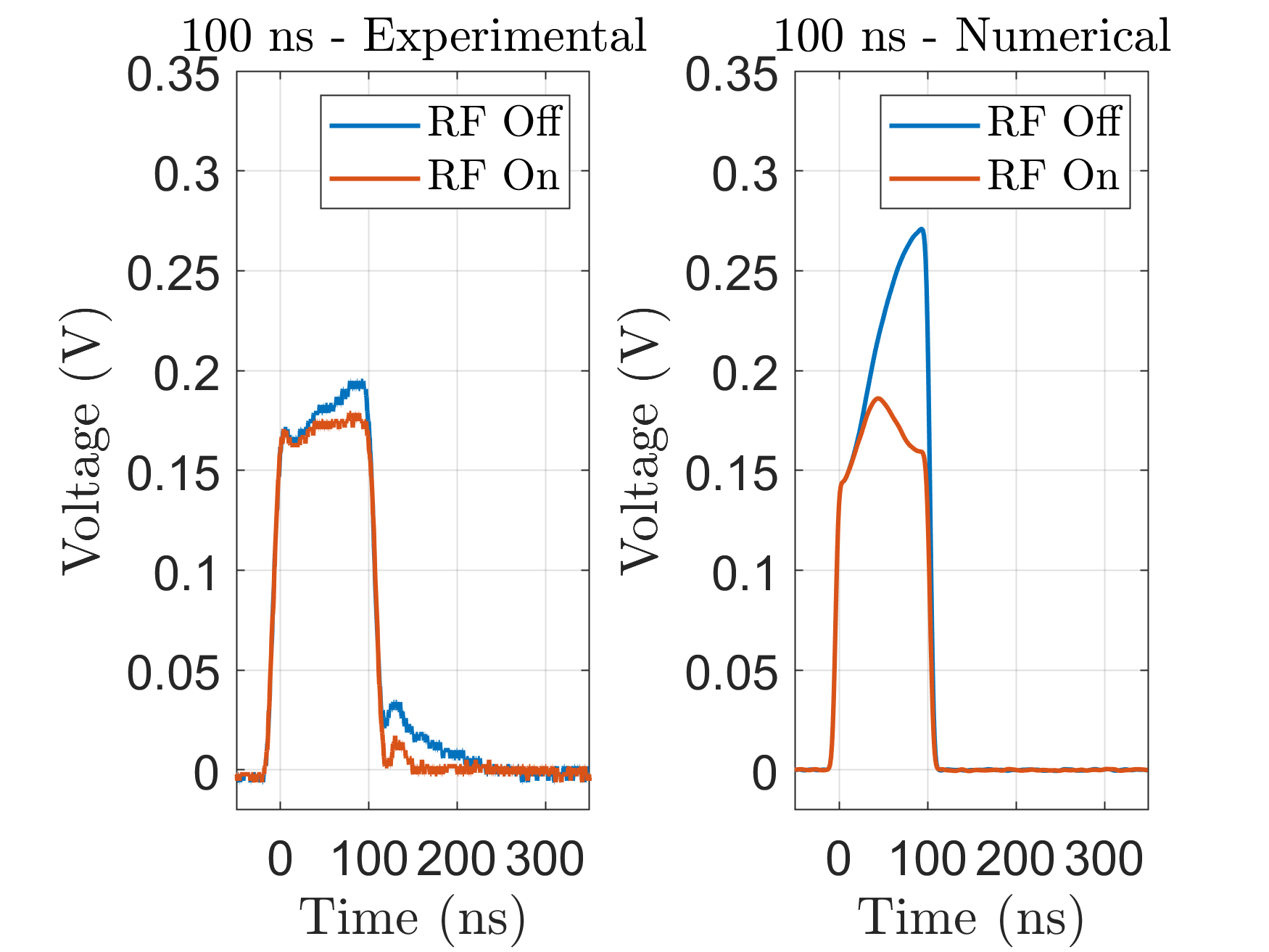}
		\label{100ns}}
	\subfloat[]{\includegraphics[width=2.5in]{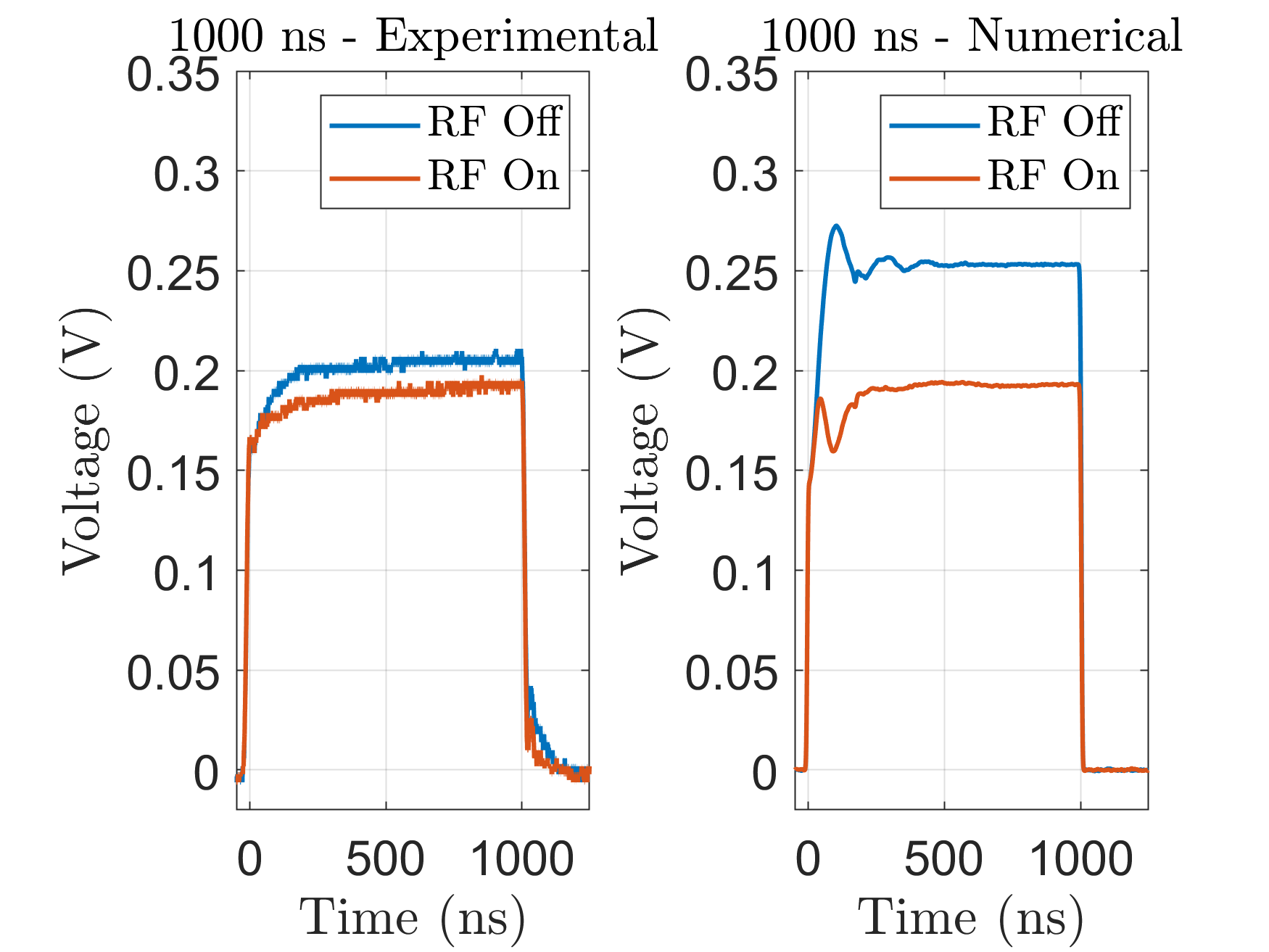}
		\label{1us}}
	\caption{Experimental and numerical transmission profiles of probe pulses after passing through Rubidium vapor. For each pulse, we plot the probe transmission with the RF source on and off. Experimental results are compared with numerical simulations for each pulse width: (a) 10ns, (b) 50ns, (c) 100ns, (d) 1000ns. Simulation results have been scaled by a factor of order one to match the scale of the experimental data.}
	\label{fig_5}
\end{figure*}

Fig 5 shows experimental measurements and simulations of the time-varying probe signal for various pulse characteristics. For the experimental data, we set the pulse widths $\tau_p$ and repetition rates $f_r$ as, $(\tau_p, 1/f_r)$ (10 ns, 500 ns), (50 ns, 1000 ns), (100 ns, 1000 ns), and (1000 ns, 2000 ns). In the simulations we only change the pulse width between trials. Comparison of the experimental results shows good agreement with theoretical predictions of the noise to within 3 dB across all pulse widths studied. The agreement of noise values indicates that physical processes in the vapor cell and detection system, are reasonably well captured by the numerical model. 

There are a few noteworthy differences between the numerical and experimental results in Fig \ref{fig_5}. First the tail at the end of each pulse observed in the experimental results is not captured numerically. This feature has been previously explored in \cite{mohl2020photon}. It is not included in the numerical model as this involves collective atomic interactions, and is not critical to the understanding of the physical processes studied in this paper. Also, the simulation underestimates the atomic response at shorter pulses. The experimental traces for 10 ns pulse show a small but clear separation while the numerical traces are almost completely overlapped. At longer pulse widths, simulations show a greater difference in transmission of the probe power with the RF on and off, when compared to experiment. This is most likely due to probe, coupling, and RF fields not being fully co-polarized in the experiment. Components of the probe beam which are not co-polarized with the RF field, will cause some atoms to not undergo AT splitting. 

Fig 5 shows a small observable difference in the measured 10 ns probe pulses when the RF source is on, versus when the RF source is off. EIT is barely established over this time scale, so little difference between an RF on/off state is observed for this RF power. For 20 ns and 50 ns probe pulses, as EIT is being actively established, distinct changes in the probe beam temporal response for RF on/off states become more apparent. For 100 ns, 200 ns, 500 ns, and 1000 ns probe pulse widths, EIT has been mostly fully established by the falling end of the probe pulses, reflected by near-constant probe voltage at the end of the pulse.

\section{Predicted Performance of Spatiotemporal Rydberg Receiver}
In this section, we utilize our experimentally benchmarked model to predict the performance of an STM Rydberg receiver for an RF communications application. We calculate the SNR as a function of probe pulse width and RF power, and then use these SNR values to predict the bit-error-ratio (BER) of the sensor for recovering an on-off-keyed (OOK) bit stream.

We define the SNR in the context of being able to detect the presence of an RF signal. For simplicity, we select an OOK RF waveform. We simulate the measured response from each probe pulse, when the RF source is on or off, as $s_{on/off,j} (t)$ for each measurement $j$. To extract the atomic response from each probe pulse, we pass $s_{on/off,j} (t)$ through a matched filter
\begin{equation}
s_{on/off,j}(\tau) = \int_{-\infty}^{\infty} d\tau\,s_{on/off,j}(t)h^*(t-\tau).
\end{equation} 
We define the matched filter $h(t)$ for a pulse of width $T$ as
\begin{equation} 
{h (t)} = 
\begin{cases}
1/T,&|t|<T/2,\\
0,&|t|\geq T/2.
\end{cases}
\end{equation}
The signal $S_j$ of measurement $j$ is then 
\begin{equation}
S_j=\text{max}[s_{on,j} (\tau)-s_{off,j} (\tau)]
\end{equation}
where the maximum takes the largest value of the difference with respect to $\tau$. The simulated error parameter $N_j$ of the receiver for measurement $j$ is defined based on simulated detector noise parameters, including shot noise, thermal noise, and dark noise based on the detector used in the experiment. $N_j$ is defined similarly to $S_j$ where voltage simulations of the signal are replaced by voltage simulations of the noise sources. Based on these definitions, the SNR of the photodetected voltage of the probe pulse is defined as
\begin{equation}
\text{SNR} = \frac{\braket{S_j}^2}{\braket{N_j^2}}
\end{equation}
Where $\braket{}$ indicates the average over repeated measurements. Computations of the SNR were made using the time-dependent 4-level model for different RF power levels and probe pulse widths. 

As Rydberg sensors make non-destructive measurements of the local E-field, we need to be careful if we are discussing a measured RF power. The typical parameters used to describe receiver antennas, such as gain and effective aperture, are not well-defined for Rydberg sensors. We choose to characterize the receiver using the incident power on the Rydberg sensor defined by the cross-sectional area of the beams through the vapor cell length times the local E-field. Given our experimental parameters of the probe beam diameter and vapor cell, our area dimensions are 0.41 mm $\times$ 7.5 cm. This area is quite small, but this is just a scaling factor to convert RF E-fields to powers.
%any practical device could use an RF field concentrating antenna.

If we assume that the RF data symbol rate is related to the probe pulse width through $RF_{\text{data rate}}=1/T$, the BER our sensor can be defined as \cite{8988790} 
\begin{equation}
\text{BER} = \frac{1}{2}\left[1-\text{erf}\left({\frac{\gamma}{2\sqrt{2}}}\right)\right] 
\end{equation}
where $\gamma$ is given as $\sqrt{\text{SNR}}$. Sometimes this quantity is referred to as the bit-error-rate, however, this is a unitless value giving the probability of measuring an error bit.  

\begin{figure}[!h]
	\centering
	\includegraphics[width=3.5in]{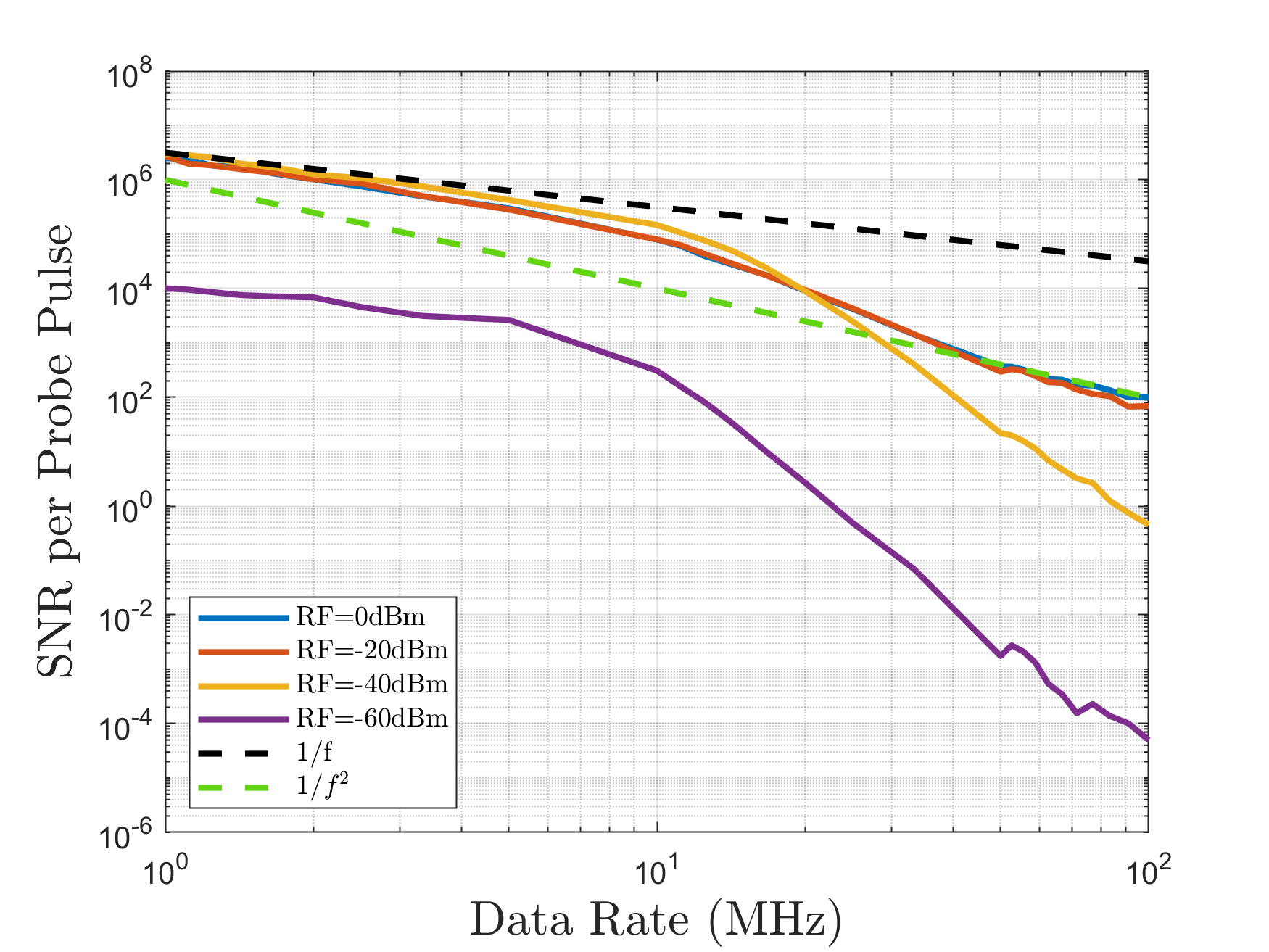}
	\caption{The SNR per probe pulse of a simulated STM Rydberg sensor as a function of RF power for different data rates. RF E-field values have been scaled by the area of the beams through the vapor cell to convert to power units. See text for more details.}
	\label{fig_6}
\end{figure}

The SNR of a single probe beam will be the same as the full STM receiver?s SNR which includes multiple probe beams. In Fig \ref{fig_6}, we plot the SNR for a range of RF powers and data rates. This figure shows that an SNR greater than 1 is achievable for 10 ns probe pulses, provided the RF power is strong enough ($>-40$ dBm) and that we only need one probe pulse per RF bit. As expected, Fig \ref{fig_6} also shows that as the RF power decreases, the SNR decreases. Additionally, as pulse width decreases, the ability to differentiate between the on and off states of the RF source, also decreases. 

In Fig \ref{fig_6}, we also plot guiding lines for the inverse of the data symbol rate $1/f$, and $1/f^2$ for strong RF fields ($>-40$ dBm).  These lines give different scaling regimes of quantum receivers. The first line, $1/f$, describes a quantum receiver with coherence time longer than a symbol period. The latter line gives the optimal scaling for a coherent quantum sensor when sensor's coherence time is shorter than a symbol period \cite{cox2018quantum}. We note that our sensor's predicted performance follows each of these lines in different regimes. For lower data rates, close to the steady state regime, our sensor follows the $1/f$ line. For higher data symbol rates, our sensor follows the $1/f^2$ line. This scaling shows promise for demonstrating a Rydberg sensor with improved SNR scaling at high data rates. The improved scaling with strong RF fields may be recoverable for weak signals by using an RF heterodyne technique.

\begin{figure}[!h]
	\centering
	\includegraphics[width=3.5in]{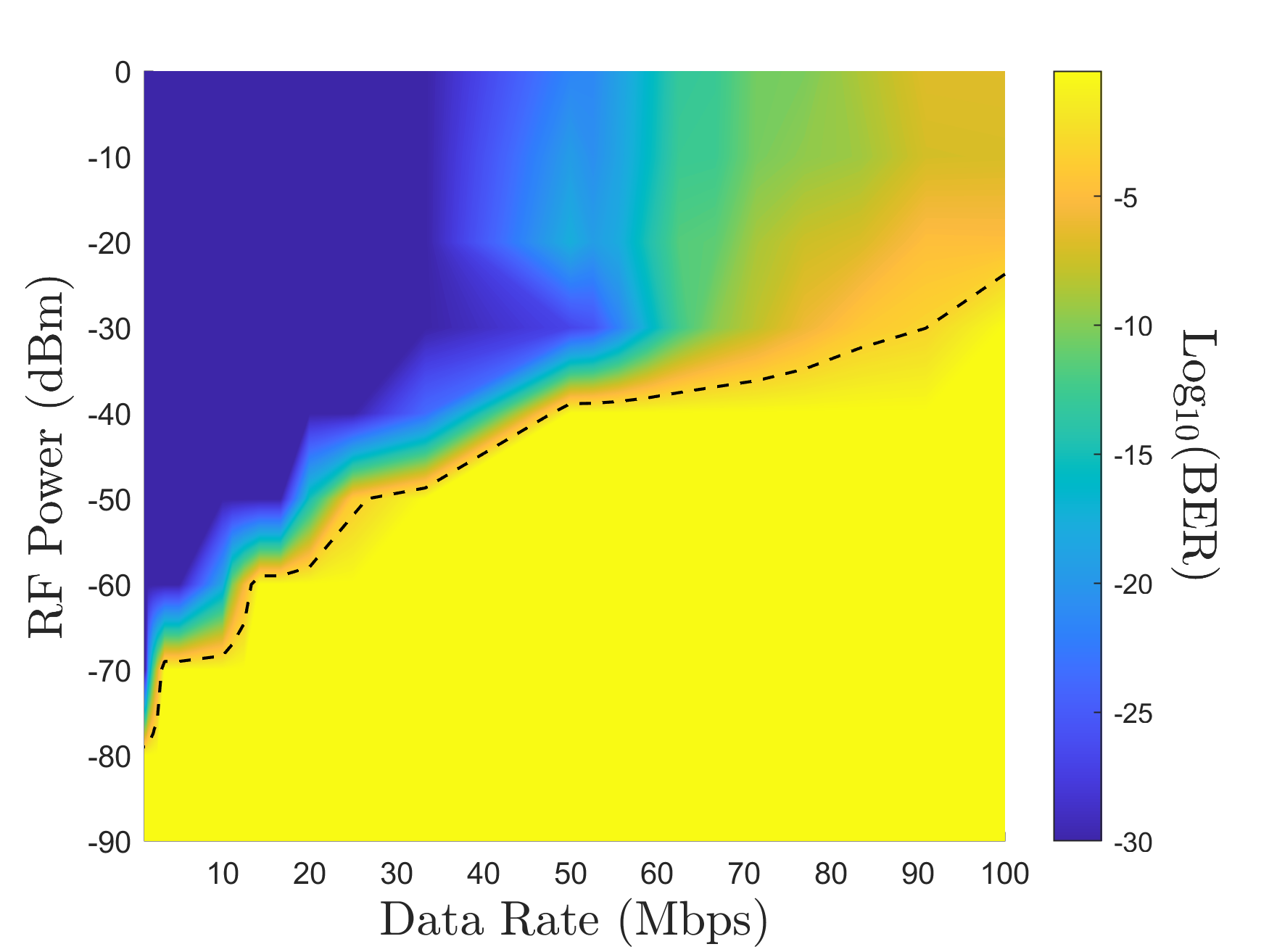}
	\caption{Predicted performance of the STM Rydberg receiver. The colormap gives the bit error ratio (BER) in $10^{-\#}$ of an OOK signal. The BER is shown as a function of RF data rate, and RF power, assuming that the RF data rate is given by $RF_{\text{data rate}}=1/T$. For visual clarity, BER values greater than -3 have been rounded to $-3$ and values less than -30 have been truncated to $30$. The black dashed line shows this transition.}
	\label{fig_7}
\end{figure}

In Fig 7, we plot a contour map of the BER on a log scale which shows that as the data rate increases, the BER increases for a fixed RF power. Additionally, as the RF power decreases, the BER also increases for a fixed data rate. To maintain error free communications with the proposed architecture at high data rates, the RF power needs to be strong enough to maintain a maximum BER on the order of  $10^{-3}$, which approximates the BER limit most forward error correction (FEC) algorithms require. Fig \ref{fig_7} shows that there is a trade between the minimum sensitivity one can achieve with the spatiotemporal receiver, and the data rate which can be measured without error. However, it is expected that there will always be a trade between data rate and RF power with this architecture, due to the build up time required to establish EIT. Nevertheless, these simulations suggest that a STM Rydberg receiver could operate at 100 MHz or higher response speeds. 

\section{Conclusion}
We have outlined and evaluated an architecture for improving the sampling rate of Rydberg atom based electric field sensors. Pulsing the probe beam allows atomic populations to reset between successive probe pulses while using the fast EIT transients to sample the RF field. Spatial and temporal multiplexing of the pulsed probe beam allow continuous sampling of an incoming RF waveform, while maintaining the improved sampling rate. RF data  rates for the STM receiver are shown to reach 100 Mbps, demonstrating that higher data rates should be possible with our architecture over current CW approaches. Other avenues for reducing the relaxation time of Rydberg based sensors are possible. For example, resonant collisional quenching, engineering of vapor cell properties, and use of different atomic vapors, could all be used in conjunction with the detailed STM receiver architecture for improving sampling rates. Further research exploring pulse shaping and sequencing for optimizing atomic population transfer may also provide a method for improving sampling rates of the STM receiver. 

\section*{Acknowledgments}
The views, opinions and/or findings expressed are those of the authors and should not be interpreted as representing the official views or policies of the Department of Defense or the U.S. Government.

{\appendix
\section*{Model}
	
	In the Rydberg receiver, the RF signal is measured using the transmission of the probe beam through the vapor cell. The time-dependent linear susceptibility of the probe beam is given by\cite{gea1995electromagnetically}
	\begin{equation}
	\chi_p(t) = \frac{2N_0d_{21}\rho_{21}(t)}{\epsilon_0E_p(t)}
	\end{equation}
	where $N_0$ is the atomic vapor density, $d_{21}$ is the dipole matrix element of the $\ket{1}\rightarrow\ket{2}$ transition, $\rho_{21}$ is the off-diagonal element of the atomic density matrix $\rho$, and $E_p$ is the electric field amplitude of the probe beam. After the linear susceptibility is defined, the time dependent absorption coefficient $\alpha(t)$ for the probe beam can be defined as $\alpha(t) = 2\pi/\lambda_p \times \text{Im}(\chi_p(t))$. The time dependent power of the probe beam as it leaves the cell, $P_p (t)$ can then be defined through a Beer-Lambert absorption law as $P_p (t)=P_i (t)\exp[-\alpha(t)L]$ where $L$ is the length of the cell, and $P_i (t)$ is the initial time dependent power of the probe beam.
	
	We model the time-dependent population dynamics of the atomic system in Fig 1 by solving the coupled system equations\cite{sandhya1997atomic}  
	\begin{equation}
	\dot{\rho} = \frac{1}{i\hbar}[H,\rho] + L(\rho).
	\end{equation}
	Assuming dipole interactions and making the usual approximations moving into the rotating frame of the electron, the Hamiltonian for the system is 
	\begin{equation}
	\begin{aligned}
	H &= \hbar\bigg[\Delta_1\ket{2}\bra{2}+\delta\ket{3}\bra{3}+\bar{\Delta}\ket{4}\bra{4}\\
	&-\frac{1}{2}(\Omega_{P}\ket{2}\bra{1}+\Omega_{C}\ket{3}\bra{2}+\Omega_{RF}\ket{4}\bra{3}+h.c)\bigg]
	\end{aligned}
	\end{equation}
	where the Rabi frequency for each beam is defined as $\Omega_j(t) = d_{lm}E_j(t)/\hbar$, and the two- and three-photon detuning are defined as $\delta = \Delta_1+\Delta_2$ and $\bar{\Delta} = \Delta_1+\Delta_2-\Delta_3$. The Lindbladian $L(\rho)$ operator captures spontaneous emission $\Gamma_{ij}$ and other dephasing processes $\gamma_{i}$ and can be written as
	\begin{equation}
	\begin{aligned}
	L(\rho) &= \frac{1}{2}\sum_{i>j}\Gamma_{ij}\left(2\sigma_{ji}\rho\sigma_{ij}-\sigma_{ii}\rho-\rho\sigma_{ii}\right)\\
	&+\frac{1}{2}\sum_{i>1}\gamma_{i}\left(2\sigma_{ii}\rho\sigma_{ii}-\sigma_{ii}\rho-\rho\sigma_{ii}\right) 
	\end{aligned}
	\end{equation}
	where $\sigma_{ij} = \ket{i}\bra{j}$. However, this does not capture transit dephasing from the motion of the atoms moving in and out of the beams. Assuming Gaussian beams with radius $w$, this dephasing can be included phenomenologically by adding to the Lindbladian\cite{meyer2021optimal}
	\begin{equation}
	L_t(\rho) = \sigma_{ii}\gamma_t - \frac{\gamma_t}{2}(I\rho+I\rho)
	\end{equation}
	with the transit dephasing rate
	\begin{equation}
	\gamma_t = \sqrt{\frac{8k_BT}{\pi m}}\frac{1}{w\sqrt{2\log(2)}}
	\end{equation}
	where $k_B$ is Boltzmann?s constant, $T$ is the cell temperature, and $m$ is the mass of Rb$^{85}$.
	
	For the Rubidium transitions listed in the main text, the dipole moments are calculated through \cite{vsibalic2017arc} giving $d_{21}=1.93ea_0,d_{32}=0.0102ea_0$, and $d_{43}=1372.2ea0$ where $e$ is the electron charge and $a_0$ is the Bohr radius. We aggregate the effects of collisional, black body, and other dephasing effects along with spontaneous emission rates into $\Gamma_{43}=\Gamma_{32}=2\pi\times 500$ kHz, and use the well known $\Gamma_{21}=2\pi\times6.066$ MHz. Our transit dephasing rate is $2\pi\times194$ kHz. Other dephasing processes are not included in the model, and so $\gamma_i=0$.
	
	\section*{Doppler Averaging}
	Doppler broadening is included in the simulations by modifying the detunings of the probe and laser beam to account for the atomic motion in the vapor cell
	\begin{equation}
	\Delta_1(v) = \omega_{21}-\omega_p-\frac{2\pi}{\lambda_p}v
	\end{equation}
	\begin{equation}
	\Delta_2(v) = \omega_{32}-\omega_c+\frac{2\pi}{\lambda_c}v
	\end{equation}
	where we have accounted for the counter-propagating beams through the sign of the atomic velocity $v$ along the optical axis.  To get the Doppler broadened results, we then velocity average using the one-dimensional Maxwell-Boltzmann distribution
	\begin{equation}
	\rho_{21,avg}(t) = \frac{1}{a\sqrt{2\pi}}\int dv\, \rho_{21}(v,t)e^{-v^2/2a^2}
	\end{equation}
	with the parameter $a=\sqrt{\frac{k_B T}{m}}$. To reduce the computational burden, we limit our simulations to velocities in the range $\pm4a$ as velocities outside this range will not significantly contribute to the Doppler average.   
}
\bibliography{IEEEabrv,IEEE_QE_STMR}
\bibliographystyle{IEEEtran}
%\newpage
\section{Biography Section}

\begin{IEEEbiographynophoto}{Samuel H. Knarr}
	received the B.S. degree in physics from Allegheny College, Meadville, PA, USA in 2012 and the Ph.D. degree in physics from the University of Rochester, Rochester, NY, USA in 2019.
	
	He is currently an optical engineer at L3Harris Technologies Inc., Palm Bay, FL, USA.
\end{IEEEbiographynophoto}

\begin{IEEEbiographynophoto}{Victor G. Bucklew}
	Victor Bucklew received the B.S. degree in Physics from the University of Madison, WI, USA in 2009 and the Ph.D. degree in electrical and computer engineering from Cornell University, Ithaca, NY, USA in 2014.  
	
	He is currently a scientist at L3Harris Technologies Inc., Palm Bay, FL, USA.
\end{IEEEbiographynophoto}

\begin{IEEEbiographynophoto}{Jerrod Langston}
	received the B.S. degree in electrical engineering from the University of Florida, Gainesville, FL, USA in 2013 and the Ph.D. degree in electrical and computer engineering from the Georgia Institute of Technology, Atlanta, GA, USA in 2019. 
	
	He is currently an electrical engineer at L3Harris Technologies Inc., Palm Bay, FL, USA.
\end{IEEEbiographynophoto}

\begin{IEEEbiography}[{\includegraphics[width=1in,height=1.25in,clip,keepaspectratio]{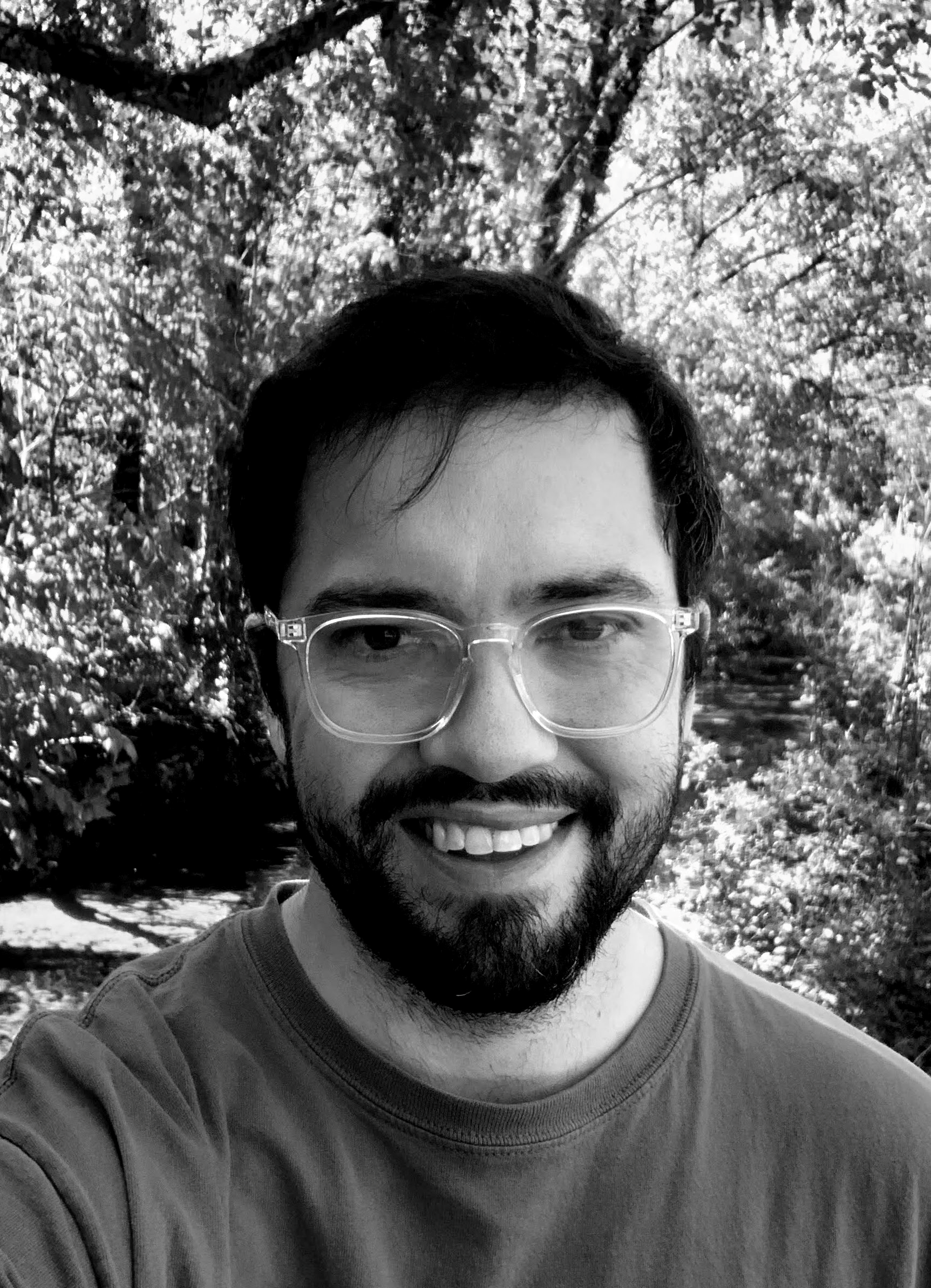}}]{Kevin C. Cox}
	received the B.S. degree in physics from The College of William and Mary, Williamsburg, VA, USA in 2011 and the Ph.D. degree in physics from the University of Colorado, Boulder, CO, USA in 2016. 
	
	Since 2016 he has been a Physicist at the U.S. Army Combat Capabilities Development Command (DEVCOM) Army Research Laboratory.  Dr. Cox's research focuses on experimental and theoretical quantum information science and quantum sensing with neutral atomic vapors.
\end{IEEEbiography}

\begin{IEEEbiographynophoto}{Joshua C. Hill}
	received the B.S degree in physics from the College of William and Mary, Williamsburg, VA, USA, and the M.S. and PhD. degrees in applied physics from Rice University, Houston, TX, USA. 
	
	He is currently a research physicist at the DEVCOM Army Research Laboratory, Adelphi, MD, USA.
\end{IEEEbiographynophoto}

\begin{IEEEbiography}[{\includegraphics[width=1in,height=1.25in,clip,keepaspectratio]{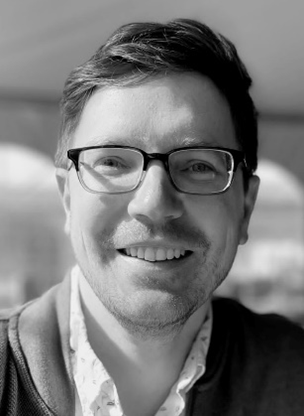}}]{David H. Meyer}
    received the B.S. degree in physics from Brigham Young University, Provo, UT, USA in 2012 and the Ph.D. degree in physics from the University of Maryland, College Park, MD USA in 2018.
	
	Since 2018 he has been a Research Physicist at the U.S. Army Combat Capabilities Development Command (DEVCOM) Army Research Laboratory. His doctoral and current research in experimental atomic physics centers on quantum information science and quantum sensing. A particular focus is developing thermal Rydberg atom-based sensors for RF electrometry and communications reception applications.
	\end{IEEEbiography}

\begin{IEEEbiography}[{\includegraphics[width=1in,height=1.25in,clip,keepaspectratio]{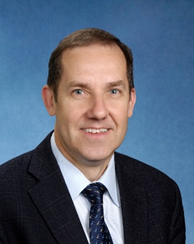}}]{James A. Drakes}
	James A. Drakes received his BS degree in Physics from the University of Dayton, Dayton, Ohio USA and M.S. and Ph.D. degrees in Physics from the University of Tennessee, Knoxville, Tennessee, USA.   
	
	With over 30 years experience, he has worked in a broad range of atomic. molecular and optical (AMO) physics domains and is currently a Senior Scientist with L3Harris Technologies focusing on quantum information science and in particular quantum sensing. 
\end{IEEEbiography}

\end{document}